\documentclass[floatfix,amsmath,amssymb,amsbsy,amsfont,aps,pre]{revtex4-2}

\usepackage{graphicx}
\usepackage{dcolumn}
\usepackage{bm}
\usepackage{caption}
\usepackage{comment}

\usepackage{subcaption}
\usepackage{xcolor}
\usepackage{titlesec}
\usepackage[colorlinks=true,linkcolor=blue]{hyperref}%

\newcommand{\figOne}{
\begin{figure}[!hbt]
    \centering
        \includegraphics[width=6.0in]{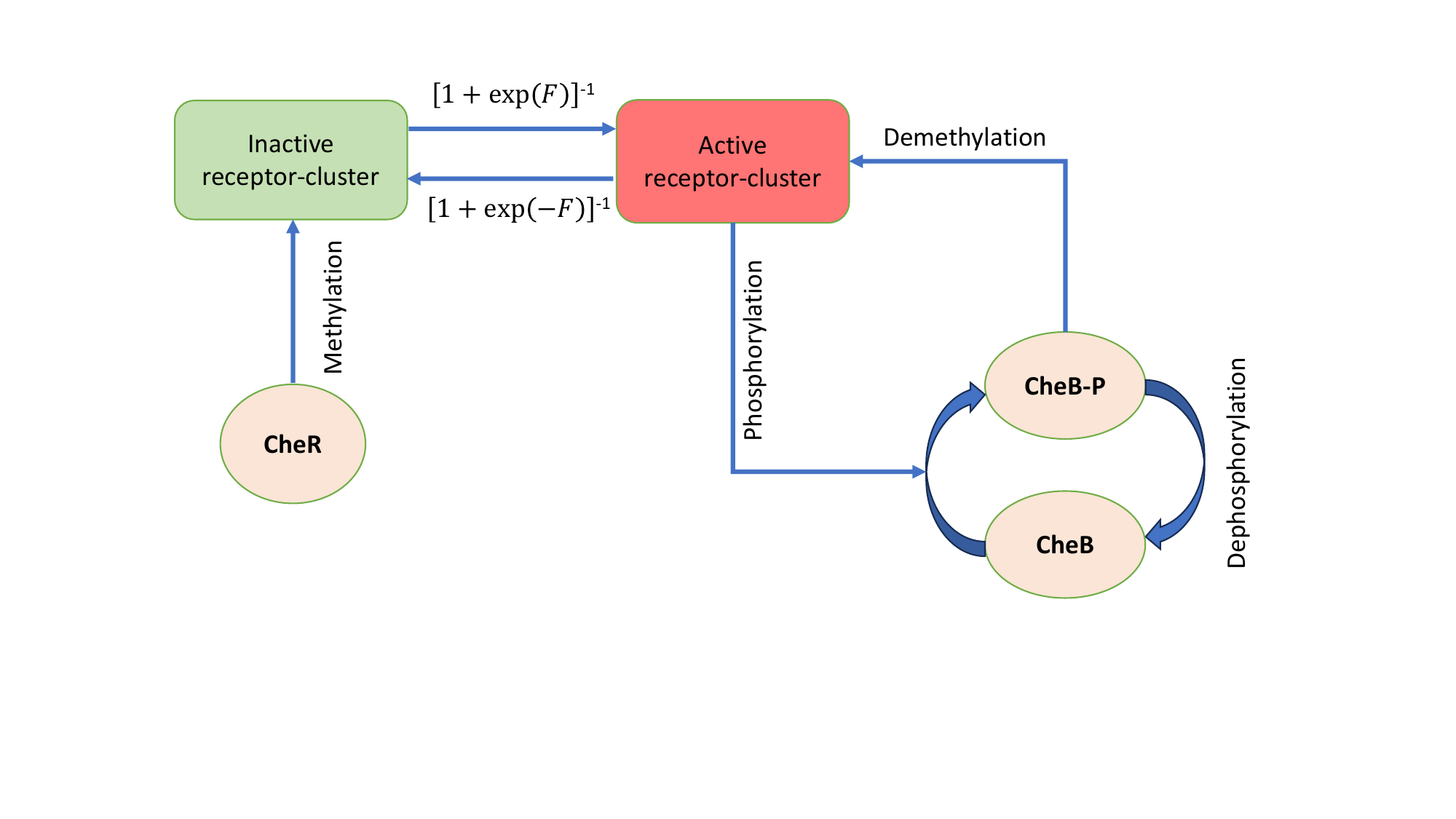}
	   \caption{Schematic representation of (de)methylation reactions and activity switching of chemoreceptor cluster.}\label{fig:1}
\end{figure}
}

\newcommand{\figTwo}{
\begin{figure*}[!hbt]
\centering
\includegraphics[width=7.0in]{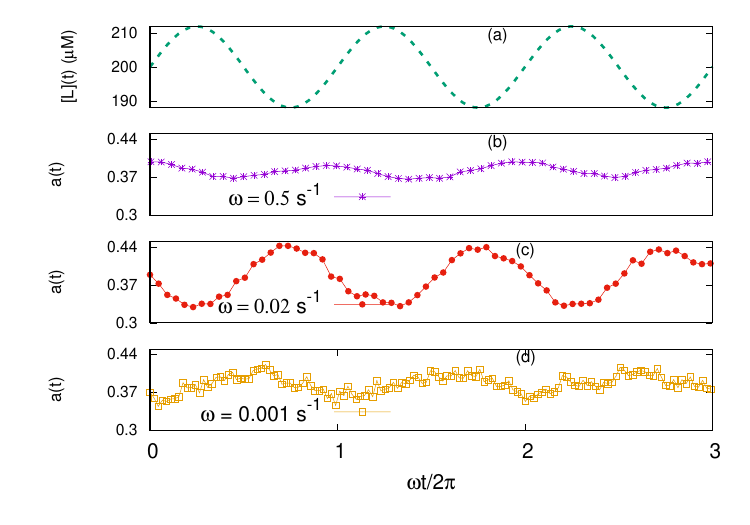}
\caption{Fraction of active receptors over three consecutive oscillation cycles measured in the steady state. Panel (a) shows the applied periodic stimulus with amplitude $[L]_1 = 10 \mu M$. Panels (b)--(d) display the corresponding activity response for frequencies  $\omega = 0.5\,\mathrm{s}^{-1}$, $\omega = 0.02\,\mathrm{s}^{-1}$, and $\omega = 0.001\,\mathrm{s}^{-1}$, respectively.  Both the amplitude and the phase lag depends on the frequency $\omega$.} \label{fig:2}
\end{figure*}
}

\newcommand{\figThree}{
\begin{figure*}[!hbt]
\centering
\includegraphics[width=7.5in]{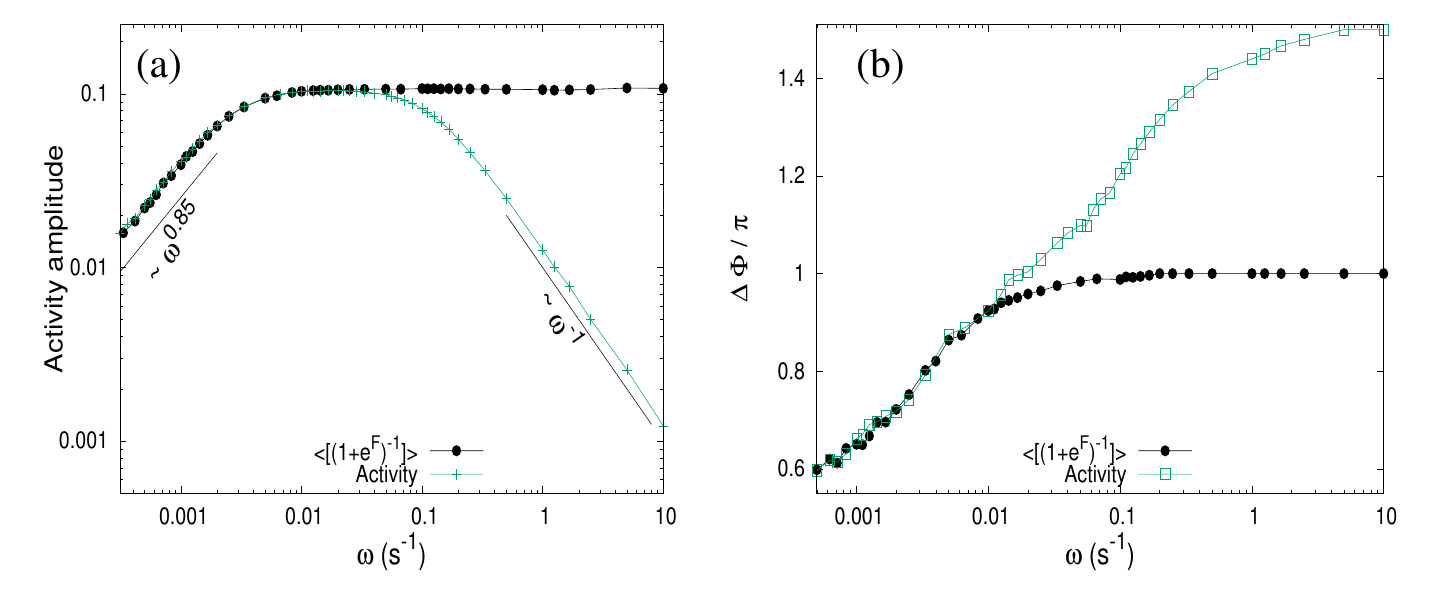}
\caption{(a) Activity amplitude (green line) as a function of applied frequency $\omega$. The amplitude increases with $\omega$, reaches a plateau and then for large $\omega$ decreases again. In the same plot we show amplitude of  $\langle [1 + \exp(F(t))]^{-1} \rangle$ (black line) which matches with activity for small and intermediate $\omega$ but deviates for large $\omega$. The maximum error-bar in these data is less than $2.5 \times 10^{-4}$. (b) Green curve shows the phase lag $\Delta \Phi$ between activity and applied stimulus as a function of $\omega$. $\Delta \Phi$ increases monotonically for small and intermediate $\omega$ and saturates at $3 \pi /2$ for large $\omega$. The black curve shows phase lag between $\langle [1 + \exp(F(t))]^{-1} \rangle$ and applied signal, which follows $\Delta \Phi$ for small and intermediate $\omega$ but saturates at $\pi$ when $\omega$ is large. The maximum error-bar in these data is less than $6 \times 10^{-3}$. All simulation parameters are as in Table \ref{table:parameters}. } \label{fig:3}
\end{figure*}
}

\newcommand{\figFour}{
\begin{figure*}[!hbt]
\centering
\includegraphics[width=6.0 in]{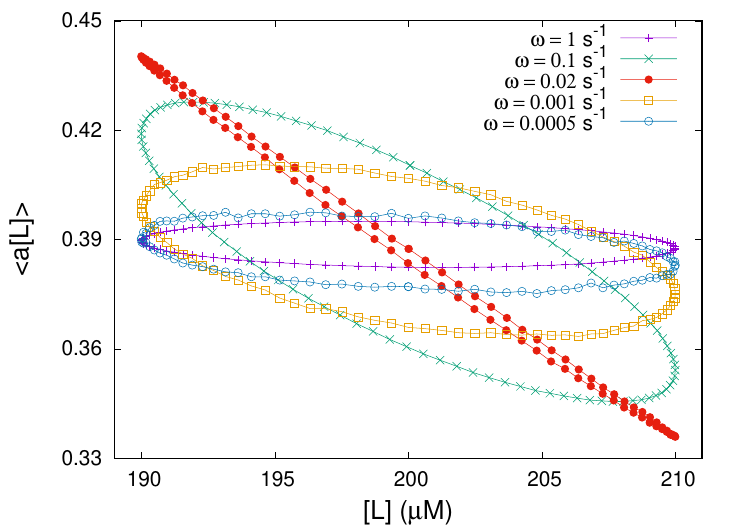}
\caption {Average activity as a function of attractant concentration over one complete time period, for different $\omega$ values. Activity traces out a loop. The area, shape and orientation of the loop depends sensitively on frequency. The statistical errors in these data points are less than the symbol size. All simulation parameters are as listed in Table~\ref{table:parameters}.} \label{fig:4}
\end{figure*}
}

\newcommand{\figFive}{
\begin{figure}[!hbt]
\centering
\includegraphics[width=7.5in]{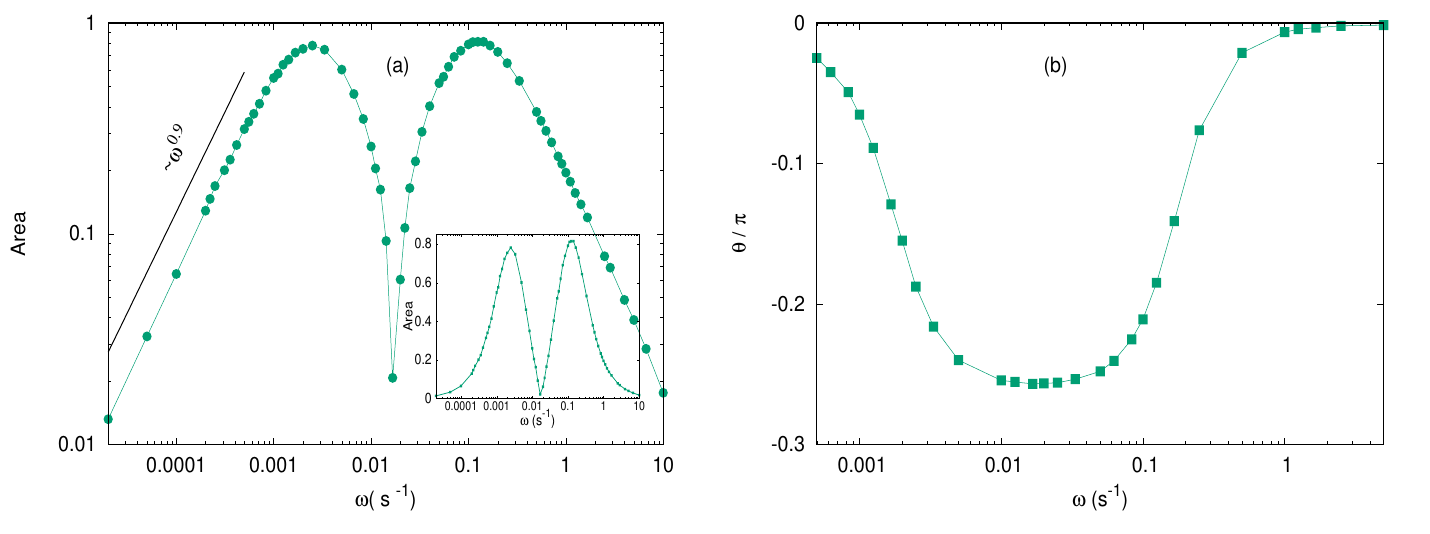}
\caption {(a) Area of the loop shows one peak at small $\omega$, another peak at large $\omega$ and decreases sharply to reach minimum for intermediate $\omega$. Main plot shows the variation in log-log scale and the inset shows it in linear scale for $y$-axis and log scale for $x$-axis. The loop area vanishes for a specific $\omega$ at which the phase lag $\Delta \Phi$ becomes equal to $\pi$. (b) The orientation $\theta$ of the loop shows a minimum with $\omega$. The error-bar is within the symbol size. All simulation parameters are listed in Table~\ref{table:parameters}. }\label{fig:5}
\end{figure}
}

\newcommand{\figSix}{
\begin{figure}[!hbt]
\centering
\includegraphics[width=6.0 in]{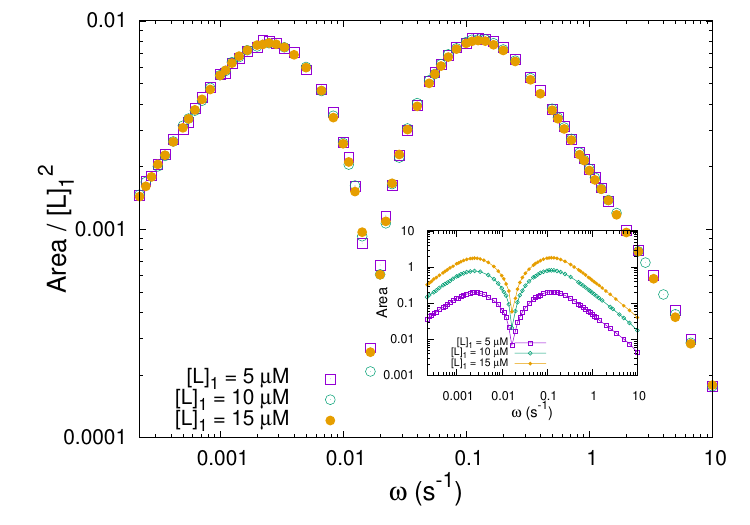}
\caption{(a) The loop area shows a scaling collapse, when scaled by square of stimulus amplitude $[L]_1^2$. Unscaled data shown in inset. The error-bar in each data point is less than the symbol size. Other simulation parameters are as in Table~\ref{table:parameters}.}\label{fig:6}
\end{figure}
}

\begin{document}
\title{Dynamics of chemo-receptor activity with time-periodic attractant field }
\author{Ramesh Pramanik, Ramu K Yadav, and Sakuntala Chatterjee}
\affiliation{Department of Physics of Complex Systems, S. N. Bose National Centre for Basic Sciences, Block JD, Sector 3, Salt Lake, Kolkata 700106, India.}
\begin{abstract}
When exposed to a time-periodic chemical signal, an {\sl E.coli} cell responds by modulating its receptor activity in a similar time-periodic manner. But there is a phase lag between the applied signal and activity response. We study the variation of activity amplitude and phase lag as a function of applied frequency $\omega$, using numerical simulations. The amplitude increases with $\omega$, reaches a plateau and then decreases again for large $\omega$. The phase lag increases monotonically with $\omega$ and finally saturates to $3 \pi /2$ when $\omega$ is large. The activity is no more a single-valued function of the attractant signal, and plotting activity vs attractant concentration over one complete time-period generates a loop. We monitor the loop area as a function of $\omega$ and find two peaks for small and large $\omega$, and a sharp minimum at intermediate $\omega$ values. We explain these results from an interplay between the time scales associated with adaptation, activity switching, and applied signal variation. In particular, for very large $\omega$ the quasi-equilibrium approximation for activity dynamics breaks down, which has not been explored in earlier studies. We perform analytical calculation in this limit and find good agreement with our simulation results.   
\end{abstract}
\maketitle

\section{Introduction} \label{sec:intro}

Bacterial chemotaxis is often studied as a model example for signal transduction network in a biological systems \cite{berg1, eisenbach, berg2, Alon, Keymer, Shimizu, Tu1, Bi, Kafri, Chatterjee}. The transmembrane chemo-receptors change their conformation upon binding with chemo-attractant molecules in the extra-cellular environment \cite{Parkinson,Yang,Hadjidemetriou}. Experiments have shown mainly two different conformations of the receptors: active and inactive \cite{Briegel,Liu}. In the active state the receptors can cause phosphorylation in the downstream network which in turn increases the tumbling bias of the cell. When the receptors are inactive, phosphorylation can not take place and the tumbling bias is low, causing the cell to swim smoothly. The fraction of active receptors is known as activity and can be directly measured in experiments \cite{Victor,Ady,Roggo}. When the cell is subjected to spatially varying or temporally varying attractant signal, it responds by changing its receptor activity and consequently its tumbling rate \cite{Chatterjee, Pramanik}. These two quantities are often measured as the response functions of the cell \cite{Shimizu, Tu1, Dev1, Dev2, Dev3, SDM1, SDM2, SDM3, Chatterjee1, Samanta1, Samanta2, Keegstra,Colin,Taejin,Frank,Zhu,Jiang,Li1}.

In many natural situations, the attractant environment shows periodic oscillations \cite{daniels2004quorum, Ishida, Tweedy}. Sometimes the attractant chemical undergoes degradation which generates a local gradient and as the cells perform chemotaxis along the gradient, a moving wavefront develops in the medium \cite{Luke,Tweedy}. In \cite{Ishida} it was shown that how pulsatile waves can be generated when attractant concentration is amplified and this information is relayed between the cells. Studying oscillatory responses is therefore biologically relevant, as E. coli in natural environments frequently encounters temporally oscillating attractant concentrations, and characterizing these responses provides insight into how the chemotactic network integrates information, adapts to fluctuations, and maintains robustness. Motivated by these questions, a number of earlier studies have considered bacterial chemotaxis in the presence of time-periodic attractant signals \cite{Shimizu, Tu1, Tu2, Zhu, Jiang, Li1, Tjalma, mandal2025bacterial}. In \cite{Shimizu,Tu1,Tu2} an oscillatory stimulus was applied and the activity response was measured. It was shown that the activity also follows a time-periodic response with the same frequency as the input signal. But there is a phase lag between the activity and the input attractant signal. Presence of a lag between the applied signal and measured response indicates the possibility of hysteresis-like behavior, when for a given value of the input, the response can have multiple values, depending on whether the rate of change of input is positive or negative, or how fast the input has been changing, etc. In the present work, we explore this aspect.

We consider an attractant profile which is spatially uniform but shows temporal oscillations with frequency $\omega$. In the long time limit the system reaches a time-periodic steady state, when the receptor activity also oscillates with the same frequency $\omega$ but with a phase delay, as mentioned in the previous paragraph \cite{Shimizu,Tu1,Tu2}. To quantify this response, we focus on two quantities, the activity amplitude, and the phase lag. Both these quantities are experimentally measurable \cite{Shimizu,Tu1,Tu2}. While the amplitude measures the gain, or the strength of the response, the phase lag reveals information about the slow adaptation dynamics. Our Monte Carlo simulations and analytical calculations reveal that the activity amplitude and the phase lag are governed by three distinct time-scales. For small $\omega$ the attractant level varies slowly and there is enough time for the methylation-demethylation network to ensure adaptation. The free energy difference between the active and inactive state of the receptor changes slowly enough such that a quasi-equilibrium state is reached and the receptor activity follows a Boltzmann distribution. As $\omega$ increases the slow methylation dynamics can not keep up with the fast changing environment anymore and the receptor free energy is controlled by the time-periodic attractant concentration now. However, the quasi-equilibrium approximation still remains valid in this regime. Finally, for very large $\omega$, the receptor activity can not switch fast enough to keep up with the rapidly changing attractant environment. Quasi-equilibrium approximation breaks down in this limit. Our analytical calculations correctly captures the variation of activity amplitude and phase and agree well with our numerics. To the best of our knowledge, this is the first ever study where the full analytical solution for time-periodic activity has been derived for a rapidly oscillating attractant environment.

Measuring the steady state activity as a function of attractant concentration over one complete time-period for a given frequency generates a closed curve or loop whose shape is found to show a sensitive dependence on the applied frequency $\omega$. We characterize the loop by its area and its orientation. The area of the loop gives information about the combined aspect of the gain and and the delay in the signal transduction system, and measures how strong the hysteresis behavior is. The loop orientation, defined as the angle between the major axis of the hysteresis loop and horizontal direction, provides a geometric aspect of the hysteresis behavior. We find the loop orientation shows a single minimum with $\omega$, but the loop area shows two maxima separated by a sharp minimum in between. To explain this, we show that at an intermediate $\omega$, the phase difference between the activity and the input attractant signal becomes exactly equal to $\pi$. This means when the attractant level passes through its minimum value, the activity touches its maximum value at exactly the same time. In this case activity becomes a single-valued function of the attractant concentration once again and there is no closed loop. We also calculate analytically the loop area for large $\omega$ and find good agreement with our simulations. Our calculations show that the loop area scales quadratically with attractant amplitude which we verify from simulations.

In the next section we describe the model. In section \ref{sec:hyst} we present our results. We make some concluding remarks, along with possible experimental verifications in Sec. \ref{sec:con}. Certain details of simulations and analytical calculations have been presented in Appendix \ref{app:tab}, \ref{app:wlarge} and \ref{app:area}.

\section{Model Description} \label{sec:model}
In Fig. \ref{fig:1} we show a schematic diagram of the signaling transduction network inside an {\sl E. coli} cell. This network consists of two parts: (A) activity switching of chemoreceptors and (B) methylation and demethylation of chemoreceptors by enzymes. The first part is responsible for sensing the chemical environment and the second part for adaptation. In this section, we discuss these two mechanisms in detail.

\subsection*{\raggedright (A) Activity Switching of Chemoreceptors}

\noindent The transmembrane chemoreceptors can be in two different states: active and inactive \cite{Briegel,Liu}. The transition between these two states are controlled by their free energy difference $F$, which is a function of the current attractant level and methylation level of the receptor. The receptors tend to switch their state cooperatively, {\sl i.e.} cluster of receptors change their state in unison. We use Monod-Wyman-Changeux (MWC) model \cite{Mello, Keymer, Monod} or `all-or-nothing' model in which each cluster contains $3n$ receptor dimers.  We have considered $n=5$ in this paper, but we have checked for $n=10,15$ our results remain unchanged. The free energy of a cluster has the form \cite{Frankel,  Dufour, Long, Pontius, Mello, Keymer, Monod}
\begin{equation}
    F = 3n \left ( 1 + \log \frac{1 + [L(t)]/K_{min}}{1 + [L(t)]/K_{max}} \right ) - \sum_{l=1}^{3n}m_l  \label{eq:MWC}
\end{equation}
where $F$ has been expressed in units of $k_BT$. In the above equation, $m_l$ denotes the methylation level of the $l^{th}$ dimer in the cluster which can take any integer value between $0$ and $8$ \cite{Pontius,Dufour,Frankel,Long}. The parameters $K_{max}$ and $K_{min}$ denote the dissociation constants of the attractant molecules from the active and inactive receptors, respectively. They also set the range of sensitivity, such that the cell can sense attractant concentration as long as it remains bounded between $K_{min}$ and $K_{max}$. In our model, we assume a spatially uniform but temporally oscillating attractant concentration, as given in Eq.~(\ref{eq:lt}). The switching rates between the active and inactive states follow local detailed balance condition \cite{Colin, SDM1, SDM2, SDM3}. An inactive cluster can switch to an active state with a rate proportional to $[1+e^F]^{-1}$, while the reverse transition rate scales as $[1 + e^{-F}]^{-1}$. 

\subsection*{\raggedright (B) Methylation and Demethylation of Chemoreceptors by Enzymes}

Apart from change in the activity state, the methylation level of the chemoreceptors also has its own dynamics. The methylation-demethylation reaction of the receptors introduces negative feedback in the signaling network which ensures adaptation. In other words, if due to sudden change in attractant concentration, the free energy $F$ in Eq. \ref{eq:MWC} changes, resulting in switching of activity state, the methylation level $m_l$ slowly changes in the opposite direction such that $F$ and activity get restored to their previous values. This is achieved by allowing only inactive receptors to get methylated and only active receptors to get demethylated. The methylating enzyme is CheR and demethylation is caused by the phosphorylated enzyme CheB-P. A CheR molecule in the cytoplasm can bind to a receptor dimer with a small rate $w_r$, provided that no other enzyme is already bound to the same dimer. Once bound, CheR increases the methylation level $m$ of the dimer by one unit with rate $k_r$, as long as the dimer belongs to an inactive receptor cluster and the methylation level has not reached the maximum ($m<8$). A bound CheR molecule can unbind with rate $w_u$, and subsequently either rebind to another unoccupied dimer in the same cluster with rate $w_{rb}$, or return to the cytoplasm. This rebinding mechanism enables a single CheR molecule to modify the methylation levels of multiple dimers within a cluster \cite{Levin, Endres, Hansen, Kim, Li}, which is essential for the robust adaptation observed in bacterial chemotaxis \cite{berg, Goy}.

The demethylating enzyme CheB undergoes phosphorylation in the presence of active receptors with rate $w_p$ and dephosphorylation with rate $w_{dp}$. The phosphorylated form, CheB-P, can bind to an unoccupied dimer with rate $w_b$. Once bound, CheB-P decreases the methylation level $m$ by one unit with rate $k_b$, provided the receptor dimer is active and $m>0$. Similar to CheR, CheB-P also undergoes unbinding and rebinding, allowing it to act on multiple receptor dimers within a cluster. The values of the kinetic parameters are summarized in Table~\ref{table:parameters}.

\figOne

\subsection*{\raggedright Time-Dependent Attractant Concentration Profile}
\noindent We consider an attractant concentration that is homogeneous in space but oscillates in time with frequency $\omega$:
\begin{equation}
 [L](t) = [L]_0 + [L]_1 \sin(\omega t). \label{eq:lt} 
\end{equation}
Here, $[L]_0$ stands for the background attractant level and $[L]_1$ represents oscillation amplitude. Throughout this work, we consider $[L]_1$ significantly smaller than $[L]_0$. Since $[L](t)$ does not depend on space, we do not explicitly include the run-tumble motion of the cell in our model. We wait for few cycles for the system to reach steady state and then measure the receptor activity. We present our results in the following section.

\section{Dynamics of Steady State Activity} \label{sec:hyst}

In Fig. \ref{fig:2} we plot a typical time series of receptor activity ({\sl i.e.} fraction of active receptors) along with the time-periodic attractant concentration, for three representative values of the input frequency $\omega$. We find that activity also shows a periodic variation in time, but there is a phase lag $\Delta \Phi$ between the activity and the input signal \cite{Shimizu,Tu1,Tu2}. Moreover, the phase lag, as well as the amplitude of activity depends on $\omega$. In Fig. \ref{fig:3} we plot the activity amplitude and the phase lag $\Delta \Phi$  as a function of $\omega$. We determine these two quantities after averaging over few oscillation cycles in steady state. We find that the amplitude increases with $\omega$, reaches a plateau and again decreases for large $\omega$. $\Delta \Phi$ varies slowly for very small $\omega$, then increases with $\omega$ and finally saturates again at $3 \pi /2$ when $\omega$ is very large. Below we discuss these variations in detail. 
\figTwo

\subsection{Frequency response of amplitude and phase}

When the input frequency $\omega$ is small, the attractant concentration varies slowly with time such that the cell can adapt to the attractant environment at all times. In this limit the receptor activity follows Boltzmann distribution with the free energy difference $F(t)$ between the active and inactive states (see Eq. \ref{eq:MWC}). In our simulations we directly measure $F(t)$ and compare the amplitude and phase lag of $\langle [1+e^{F(t)}]^{-1}\rangle $ with those of receptor activity in Figs. \ref{fig:3}A and \ref{fig:3}B, respectively. Here, the angular brackets denote the averaging performed over steady state ensemble. At small frequency, we indeed find very good agreement. In this quasi-equilibrium limit, an approximate analytical calculation of activity was performed in \cite{Shimizu,Tu1,Tu2} and it was shown that the amplitude increases linearly with $\omega$. In our simulations, we observe a slightly different power law growth, $\sim \omega^{0.85}$. Instead of using a mean field description for methylation dynamics, as done in \cite{Shimizu,Tu1,Tu2}, we have explicitly modeled the binding-unbinding kinetics of the enzymes (see Sec. \ref{sec:model}) \cite{SDM1,SDM2,SDM3,Chatterjee1,Colin}. The difference in the power law exponent may be because of this difference in the methylation dynamics description. In \cite{Shimizu,Tu1,Tu2} phase lag was also calculated and it was shown that for $\omega \to 0$ phase lag is $\pi /2$. Numerically, it was difficult for us to access such small frequency regime and hence the phase lag we observe at lowest $\omega$ value is still larger than $\pi /2$.

As the input frequency $\omega$ becomes larger, the slow methylation-demethylation reactions cannot keep up with the rapidly changing attractant environment. The variation in $F(t)$ is then fully controlled by $[L](t)$. The activity is still given by $\langle [1+e^{F(t)}]^{-1} \rangle $ and its amplitude can be estimated from finding which value of $[L](t)$ minimizes $F(t)$. It is easy to see from Eq. \ref{eq:lt} that $F(t)$ is minimum when $\sin (\omega t)$ is minimum, {\sl i.e.} for $\omega t = 3 \pi /2$, when $[L](t) = [L]_0 - [L]_1 $. The phase lag between the activity and ligand profile is therefore $3 \pi /2 - \pi /2 =\pi$. Since the methylation level hardly changes here, the activity amplitude does not depend on $\omega$ in this range. However, as $\omega$ increases further and becomes comparable or larger than the rate $w_a$ at which the receptor clusters switch their activity, the quasi equilibrium approximation breaks down. The activity is not given by $\langle [1+e^{F(t)}]^{-1} \rangle $ anymore. In this case the time-evolution of activity follows
\begin{equation}
\frac{d\langle a(t) \rangle}{dt} =\frac{w_a}{1+e^{F(t)}} [1-\langle a(t) \rangle ] -\frac{w_a e^{F(t)}}{1+e^{F(t)}} \langle a(t) \rangle, \label{eq:dynamics} 
\end{equation}
where $\langle a(t) \rangle$ denotes the average fraction of active receptors at time $t$. Since methylation remains effectively constant in this fast time-scale, time-dependence of $F(t)$ is solely governed by $[L](t)$. Above equation can be solved analytically (see Appendix \ref{app:wlarge} for details) and in the long time limit, we have 
\begin{equation}
\langle a(t) \rangle \approx \frac{C_1}{w_a} + \frac{C_2}{\omega} \cos (\omega t ), 
\label{eq:act}
\end{equation}
where $C_1$ and $C_2$ are $\omega$-independent constants whose detailed expressions are given in Appendix \ref{app:wlarge}. In Eq. \ref{eq:act} the coefficient $C_1$ determines the dc value of activity and depends only on the background attractant level $[L]_0$. The coefficient $C_2$ appears in the time-periodic part of activity response and depends on the ratio $[L]_1/[L]_0$. It also follows from Eq. \ref{eq:act} that the amplitude of $\langle a(t) \rangle $ scales as $1/\omega$, which agrees with our simulation results in Fig. \ref{fig:3}A. Since $\langle a(t) \rangle $ lags behind $[L](t)$, its peak of $\langle a(t) \rangle$ which occurs after the peak of $[L](t)$ at $\omega t = \pi /2$, can be found at $\omega t = 2 \pi$, which means the phase lag takes the value $2 \pi - \pi /2 = 3 \pi /2$, which is consistent with our simulation results in \ref{fig:3}B. 
\figThree

\subsection{Activity traces out a loop with attractant signal}

We measure $\langle a(t) \rangle$ as a function of $[L](t)$ during one complete cycle in steady state. This generates a loop, which we show in Fig. \ref{fig:4} for different frequencies. Depending on the frequency, the shape and size of the loop vary. To characterize this variation quantitatively, we plot the area of the loop as a function of frequency in Fig. \ref{fig:5}A. We use trapezoidal rule to calculate the loop area. We find the loop area shows two maxima and one in-between minimum as frequency is varied. For small $\omega$, the loop area increases as a power law with exponent $\sim 0.9$ whereas for large $\omega$ it decreases as $1/\omega$. We analytically calculate the loop area for large $\omega$ in Appendix \ref{app:area} and find that the area equals $[L]_1 C_2 \pi /\omega$. This matches with our simulation observation. Moreover, after substituting the form for $C_2$ it follows that for a fixed $\omega$ area scales as $[L]_1^2$. Although this result is derived only for large $\omega$, our simulations in Fig. \ref{fig:6} show that for all $\omega$ values area $\sim [L]_1^2$ scaling remains valid.
\figFour

Fig. \ref{fig:5}A also shows that the loop area dips sharply at intermediate frequencies. This means the loops get very narrow, which can also be seen from our data in Fig. \ref{fig:4}. This phenomenon can be explained from the fact that the phase lag between activity and attractant signal is close to $\pi$ in this frequency range (see Fig. \ref{fig:3}B). A $\pi$ phase lag means the maximum (minimum) of $[L](t)$ and minimum (maximum) of $\langle a(t) \rangle$ occur together. As $[L](t)$ decreases (increases) from its maximum (minimum), activity also retraces its steps. This means for every value of $[L](t)$ there is a unique value of $\langle a(t) \rangle$, and therefore no closed loop. We find this effect for $\omega \sim 0.02s^{-1}$. For $\omega$ close to this value, although we find a loop but its area remains small. 
\figFive

In addition to the loop area, the loop orientation varies with applied frequency Fig. \ref{fig:4}. We define $\theta$ as the clockwise angle between the loop’s major axis—obtained via principal component analysis—and the horizontal axis. As shown in Fig. \ref{fig:5}B, $\theta$ decreases with $\omega$, reaches a minimum, and then rises toward zero at high frequencies. We do not have a simple explanation behind this behavior.

\figSix

\section{Conclusions}
\label{sec:con}

In this work we have studied the activity response of chemo-receptors in an {\sl E.coli} cell which is exposed to a stimulus whose level shows time-periodic oscillations. While previous studies \cite{Shimizu,Tu1,Tu2} reported presence of a phase lag between the receptor activity and applied attractant signal, the hysteresis-like effects caused by this phase lag were not explored earlier, to the best of our knowledge. We show here that the loop area shows an interesting frequency response, reaching maximum values for low and large frequencies and dropping to zero for intermediate frequency when phase lag equals $\pi$. The biochemical pathway model that we use, accounts for the fast but finite switching rate of receptor activity \cite{Colin, SDM1, SDM2, SDM3, Chatterjee1}. When the applied frequency becomes larger than the activity switching time-scale, the signal varies too fast for the system to respond. We have been able to perform analytical calculations in this regime and our calculations quantitatively explains the variation of activity amplitude, phase lag and the loop area for large $\omega$. This regime remained unexplored in previous studies \cite{Shimizu,Tu1,Tu2} where it was assumed receptors change between active and inactive conformations infinitely fast.

Our study brings out the interplay between three different time-scales in the system: activity switching time-scale $\tau_a$, methylation time-scale $\tau_m$ and attractant oscillation time-scale $1/\omega$. The signaling network always satisfies $\tau_a \ll \tau_m$, but different behavior is observed depending on whether $1/\omega$ falls above or below these two time-scales or has an intermediate value \cite{Tjalma}. We present our main findings in a concise manner in Table \ref{table:summarize}.

\begin{table}[htbp!]
\centering
\caption{Activity amplitude, phase and loop area in different frequency domains}
\begin{tabular}{|p{2.8cm}| p{3.6cm}| p{2.8cm}| p{2.5cm}| p{3.2cm}|}
\hline
\textbf{{Frequency regime and order of time-scales}} & \textbf{Methylation and free energy} & \textbf{Activity amplitude} & \textbf{Phase lag between input signal and output} & \textbf{Loop area} \\
\hline
{Low frequency}: \newline $\tau_a \ll \tau_m \ll 1/\omega$ & Methylation keeps up with attractant variation, system remains adapted, quasi-equilibrium approximation works. &  goes to $0$, as $\omega \to 0$. &  $ \pi/2$. & goes to $0$, as $\omega \to 0$.  \\[0.6em]
\hline
{Intermediate frequency}: \newline $\tau_a \ll 1/\omega \ll \tau_m$ & Methylation cannot change fast enough and $F$ is controlled by attractant variation; quasi-equilibrium approximation remains valid. & stays constant at its maximum value. & increases with $\omega$ from $ \pi/2$ to  $ 3\pi/2$. & shows two maxima and an intermediate minimum with $\omega$. \\[0.6em]
\hline
{High frequency}: \newline $1/\omega \ll \tau_a \ll \tau_m $ & Methylation stays almost constant; quasi-equilibrium approximation breaks down. & goes to $0$, as $\omega \to \infty.$ & $ 3\pi/2$. & vanishes as $\omega \to \infty$.  \\[0.6em]
\hline
\end{tabular}
\label{table:summarize}
\end{table}

Our main conclusions should not depend on the microscopic details of our model. For example, the specific way in which we model binding-unbinding events between the enzyme molecules and receptor dimers, is not expected to have any major qualitative effect on the frequency response of activity dynamics. Therefore, our conclusions should remain valid even if a different model is used for the biochemical pathway. Also, the receptor activity directly controls the tumbling bias of the cell \cite{SDM1, SDM2, Chatterjee1}. Just like activity, a closed loop is also expected when tumbling bias is plotted against attractant level over one complete cycle.

It should be possible to experimentally test the main results using established FRET (fluorescence resonance energy transfer) based measurement techniques. Previously, some studies were done where FRET microscopy was employed to monitor kinase activity in E. coli cells \cite{Shimizu, Frank, Tu1, Tu2}. In such experiments, cells were immobilized on coated glass surfaces and placed inside a chamber that allowed precise control of the surrounding chemical environment. Fluorescent signals were used to monitor interactions between intracellular molecules in response to changes in the chemical conditions. These measurements made it possible to track changes in the cell’s kinase activity. Using time-periodic stimulus the frequency response of amplitude and phase lag of activity was already measured in \cite{Shimizu} over a certain frequency range. Similar set-ups can be used to test our conclusions.

Finally, we have considered a simple sinusoidal time-variation in this work. But a time-periodic signal can in general have a more complex form, like superposition of sine waves with different frequencies, or an attractant profile with periodicity both in space and time as in a traveling wave environment \cite{mandal2025bacterial}. It might be of interest to explore how the interplay between the adaptation time-scale, activity switching time-scale and multiple time-scales present in the input signal affects the response dynamics of the system.

\section{Acknowledgements} 

RP acknowledges the research fellowship (Grant No. 09/0575(13013)/2022-EMR-I) from the Council of Scientific and Industrial Research (CSIR), India. SC acknowledges support from the Anusandhan National Research Foundation (ANRF), India (Grant No. CRG/2023/000159).

\appendix

\section{Data Availability Statement and Author Contribution}
All relevant data are submitted electronically as supporting information. Ramesh Pramanik carried out the simulations, data analysis, and analytical calculations, and prepared the figures. Ramu K. Yadav contributed to the simulations and data analysis. Sakuntala Chatterjee designed and supervised the project, interpreted the results, and wrote the manuscript. All authors reviewed and approved the final version of the manuscript. 

\section{List of parameter values used in simulations}
\label{app:tab}

In this section, we provide all simulation parameters.
\begin{table}[htbp!]
\centering
\caption{Values of simulation parameters}
\label{table:parameters}
\begin{tabular}{|l|l|l|l|} 
 \hline
 Symbol & \hspace{1 pt} Description & Value & References  \\ [0.9ex] 
 \hline
$N_{dim} $ & Total number of receptor dimers & 7200  &  \cite{Pontius, Li}\\ 
\hline
$n$ & Number of trimers of dimers in a receptor cluster & 5 & Present study \\
\hline
$[L]_0$ & Background concentration level & 200 $\mu$M &  \cite{SDM1,SDM2,SDM3,Chatterjee1} \\
\hline
$[L]_1$ & Amplitude of oscillating concentration profile & 10 $\mu$M &  Present study \\
\hline
$K_{max}$ & Ligand dissociation constant for active receptor & 3200 $\mu$M & \cite {Flores,Jiang} \\
\hline
$K_{min}$ & Ligand dissociation constant for inactive receptor & 7 $\mu M$  & \cite {Flores,Jiang} \\ 
\hline
$N_R$ & Total number of CheR enzyme & 140 &  \cite{Pontius, Li}  \\
\hline
$N_B$ & Total number of CheB enzyme & 240  &  \cite{Pontius, Li} \\
\hline
$w_r$ & Binding rate of bulk CheR to tether site of an unoccupied dimer & 0.068 $s^{-1} $ & \cite{Pontius, Sonja}  \\
\hline
$w_b$ &  Binding rate of bulk CheB-P to tether site of an unoccupied dimer & 0.061 $s^{-1}$   & \cite{Pontius, Sonja} \\
\hline
$w_u$ & Unbinding rate of bound CheR and CheB-P enzyme & 5 $s^{-1}$   & \cite{Pontius, Sonja}  \\
\hline
$k_r$ & Methylation rate of a dimer by a bound CheR protein & 2.7 $s^{-1}$  & \cite{Pontius, Sonja}  \\
\hline
$k_b$ & Demethylation rate of a dimer by a bound CheB-P protein &3 $s^{-1}$  & \cite{Pontius, Sonja}  \\
\hline
$w_p$ & Phosphorylation rate of a CheB enzyme & 3 $s^{-1}$  & \cite{Pontius, Stewart}\\
\hline
$w_{dp}$ &  Dephosphorylation rate of a CheB-P enzyme & 0.37 $s^{-1}$  & \cite{Pontius} \\
\hline
$w_{rb}$ & Rebinding rate of CheR and CheB-P & 1000 $s^{-1}$ & \cite{Pontius} \\
\hline
$w_a$ & Switching rate of activity & 0.25 $s^{-1}$  & \cite{Colin} \\
\hline
$dt$ & Time step & 0.01 s  & \cite{SDM1,SDM2,SDM3,Chatterjee1} \\
\hline
\end{tabular} 
\end{table}

\section{Analytical calculation in large {$\omega$} limit} \label{app:wlarge}

The fraction of active receptors at time $t$ is denoted as $a(t)$, which is a fluctuating quantity, as shown in Fig. \ref{fig:2}. The mean activity $\langle a(t) \rangle$ is obtained after averaging over a large number of cycles. In this appendix we calculate $\langle a(t) \rangle$ in the limit of large frequency. The time-evolution equation for $\langle a(t) \rangle$ can be written in terms of transitions between active and inactive states of the receptor cluster. Using Eq.~(\ref{eq:dynamics}), we obtain
\begin{equation}
    \frac{d\langle a(t) \rangle}{dt} + w_a\langle a(t) \rangle = \frac{w_a}{1+ \exp( F(t))}.
    \label{eq:at}
\end{equation}
The expression for free energy $F(t)$ is given in Eq. 1 and we further simplify it as
\begin{equation}
     F(t) = 3n \log \frac{1 + [L(t)]/K_{min}}{1 + [L(t)]/K_{max}}  - 3nm,
    \label{eq:free}
\end{equation} 
where $m$ stands for average methylation level of a receptor dimer within the cluster. We estimate its value towards the end of this appendix. The form for time-varying ligand concentration $[L](t)$ has been given in Eq. 2 in main text. To ensure that the system is far from the limits of sensitivity, we choose  $[L](t)$ such that $K_{max} >> [L(t)] >> K_{min}$ for all $t$. This simplifies $F(t)$ further as $F(t) \approx 3n \left ( \log \dfrac{[L](t)}{K_{min}}-m \right )$. Now, when $\omega$ is large, $[L](t)$ varies rapidly but the receptor methylation level can not keep up since the time-scale associated with methylation dynamics is slow (see low values of binding rate constants of the enzymes in Table~\ref{table:parameters}). Therefore, for large $\omega$ we can assume that $m$ remains almost constant with time. Eq. \ref{eq:at} then becomes 
\begin{equation}
    \frac{d\langle a(t) \rangle}{dt} + w_a\langle a(t) \rangle = C_1  - C_2 \sin(\omega t),
     \label{eq:c1c2}
\end{equation}
where we have defined
\begin{equation}
C_1 = \frac{w_a}{1 + B^{3n}}
\end{equation}
and
\begin{equation}
\frac{C_2}{3nw_a} = \frac{[L]_1}{[L]_0} \frac{B^{3n}}{[1+B^{3n}]^2} \label{eq:c2}
\end{equation}
with $B = e^{-m}[L]_0 /K_{min}$. Here, we have also used the fact that $[L]_1$ is significantly less than $ [L]_0$ and retained terms only up to linear order in $\dfrac{[L]_1}{[L]_0}$. The solution to Eq. \ref{eq:c1c2} in the long time limit looks like
\begin{equation}
    \langle a(t) \rangle = \frac{C_1}{w_a}  + \frac{C_2 w_a}{\omega^2 + w_a^2} sin(\omega t)  + \frac{C_2 \omega}{\omega^2 + w_a^2} cos(\omega t).
\end{equation}
For $\omega \gg w_a$
\begin{equation}
\langle a(t) \rangle = \frac{C_1}{w_a} + \frac{C_2}{\omega} \cos(\omega t)
\label{eq:soln}
\end{equation}
which is same as Eq. 4 in main text. The coefficients $C_1$, $C_2$ depend on $m$, the average methylation level of a receptor dimer. We estimate $m$ from our simulation data, using the linear fit of activity amplitude vs $\omega$ in large $\omega$ range. We find $m \simeq 3.17$.

\section{Analytical calculation of loop area for large {$\omega$} }
\label{app:area}

The loop is obtained by plotting the variation of receptor activity as a function of attractant concentration over one complete cycle. The area of the loop is defined as 
\begin{equation}
\text{Area} = \oint \langle a[L(t)] \rangle d([L(t)]),
\end{equation}
where the integration is performed over one time period of variation. Above equation can be rewritten as 
\begin{equation}
\text{Area} = \int_0^{2 \pi /\omega} \langle a(t) \rangle \frac{d [L](t)}{dt} dt
\end{equation}
Using Eq. \ref{eq:soln} and Eq. 2 from main text
\begin{equation}
    \text{Area} = \int_{0}^ {2 \pi /\omega} \left ( \frac{C_1}{w_a} + \frac{C_2}{\omega} \cos(\omega t) \right ) [L]_1\omega \cos(\omega t) dt  = \frac{[L_1]C_2\pi}{\omega}.
     \label{eq:area}
\end{equation}
Substituting $C_2$ from Eq. \ref{eq:c2} the scaling form shown in Fig. 6 can be readily verified.

\end{document}